\documentclass[10pt,A4paper,twocolumn,aps,pra, superscriptaddress,longbibliography, nofootinbib]{revtex4-2}

\usepackage{amsmath}    
\usepackage{graphicx}   
\usepackage{color}      
\usepackage[usenames,dvipsnames,svgnames,table]{xcolor}
\usepackage[hidelinks]{hyperref}   
\usepackage{braket}
\usepackage{bbm}

\usepackage[capitalise]{cleveref}
\Crefname{figure}{Fig.}{Figs.}
\Crefname{table}{Tab.}{Tabs.}
\usepackage{siunitx}
\usepackage{amssymb}

\usepackage{rotating}
\usepackage{tabularx}
\usepackage{cellspace}
\usepackage{booktabs}
\usepackage{graphics}
\usepackage{newfloat}
\usepackage{float}
\newcolumntype{Y}{>{\centering\arraybackslash}X}
\newcolumntype{d}{c}
\newcolumntype{a}{c}

\usepackage{xr}

\newcommand{\trace}{\ensuremath{\mathrm{Tr}}}

\newcommand{\g}{\ensuremath{\ket{g}}}
\newcommand{\e}{\ensuremath{\ket{e}}}
\newcommand{\f}{\ensuremath{\ket{f}}}

\newcommand{\ns}[1]{\ensuremath{\SI{#1}{\nano\second}}}
\newcommand{\mus}[1]{\ensuremath{\SI{#1}{\micro\second}}}
\newcommand{\mhz}[1]{\ensuremath{\SI{#1}{\mega\hertz}}}
\newcommand{\tlru}{\ensuremath{\tau_\mathrm{LRU}}}
\newcommand{\modamp}{\ensuremath{\omega_\mathrm{a}}}
\newcommand{\modampflux}{\ensuremath{\phi_\mathrm{a}}}
\newcommand{\modfreq}{\ensuremath{\omega_\mathrm{m}}}

\newcommand{\oppoint}{\ensuremath{O}}
\newcommand{\tsymbol}{\ensuremath{t}}
\newcommand{\durationsymbol}{\ensuremath{\tau}}
\newcommand{\bufferduration}{\ensuremath{\tau_\mathrm{B}}}
\newcommand{\fr}{\ensuremath{\omega_{\mathrm{r}}}}
\newcommand{\frb}{\ensuremath{\omega_{\mathrm{r, b}}}}
\newcommand{\gb}{\ensuremath{g_{\mathrm{qr}}}}
\newcommand{\fp}{\ensuremath{\omega_{\mathrm{p}}}}
\newcommand{\kr}{\ensuremath{\kappa_\mathrm{r}}}
\newcommand{\kp}{\ensuremath{\kappa_\mathrm{p}}}
\newcommand{\ejmax}{\ensuremath{E_J^{\mathrm{max}}}}

\newcommand{\fge}[1][\empty]{%
  \ifx#1\empty
    \ensuremath{\omega_\mathrm{ge}}
  \else
    \ensuremath{\overline{\omega}_\mathrm{ge}}
  \fi
}
\newcommand{\fef}[1][\empty]{%
  \ifx#1\empty
    \ensuremath{\omega_\mathrm{ef}}
  \else
    \ensuremath{\overline{\omega}_\mathrm{ef}}
  \fi
}

\newcommand{\rtime}{t}

\newcommand{\modfreqvalue}{564}
\newcommand{\modampvalue}{128}

\newcommand{\puttitle}{Fast Flux-Activated Leakage Reduction for Superconducting Quantum Circuits}

\def \hrho{\hat{\rho}}
\def \hphit{\hat{\varphi_\mathrm{t}}}
\def \hnt{\hat{n}_\mathrm{t}}
\def \hf{\hat{f}}
\def \ha{\hat{a}}

\def \hH{\hat{H}}

\newif\ifshowsectionnames
\newcommand{\mysection}[1]{\ifshowsectionnames\section{#1}\fi}
\showsectionnamesfalse

\begin{document}

\title{\puttitle}

\author{Nathan Lacroix}
\affiliation{Department of Physics, ETH Zurich, CH-8093 Zurich, Switzerland}

\author{Luca Hofele}
\affiliation{Department of Physics, ETH Zurich, CH-8093 Zurich, Switzerland}

\author{Ants Remm}
\affiliation{Department of Physics, ETH Zurich, CH-8093 Zurich, Switzerland}

\author{Othmane Benhayoune-Khadraoui}
\author{Alexander McDonald}
\author{Ross Shillito}
\affiliation{Institut Quantique and Département de Physique, Université de Sherbrooke, Sherbrooke J1K2R1 Québec, Canada}

\author{Stefania Lazar}
\affiliation{Department of Physics, ETH Zurich, CH-8093 Zurich, Switzerland}

\author{Christoph Hellings}
\affiliation{Department of Physics, ETH Zurich, CH-8093 Zurich, Switzerland}

\author{Francois Swiadek}
\affiliation{Department of Physics, ETH Zurich, CH-8093 Zurich, Switzerland}

\author{Dante Colao-Zanuz}
\affiliation{Department of Physics, ETH Zurich, CH-8093 Zurich, Switzerland}

\author{Alexander Flasby}
\affiliation{Department of Physics, ETH Zurich, CH-8093 Zurich, Switzerland}
\affiliation{ETH Zurich - PSI Quantum Computing Hub, Paul Scherrer Institute,CH-5232 Villigen, Switzerland}
\author{Mohsen Bahrami Panah}
\affiliation{Department of Physics, ETH Zurich, CH-8093 Zurich, Switzerland}
\affiliation{ETH Zurich - PSI Quantum Computing Hub, Paul Scherrer Institute,CH-5232 Villigen, Switzerland}
\author{Michael Kerschbaum}
\affiliation{Department of Physics, ETH Zurich, CH-8093 Zurich, Switzerland}
\affiliation{ETH Zurich - PSI Quantum Computing Hub, Paul Scherrer Institute,CH-5232 Villigen, Switzerland}
\author{Graham J. Norris}
\affiliation{Department of Physics, ETH Zurich, CH-8093 Zurich, Switzerland}

\author{Alexandre Blais}
\affiliation{Institut Quantique and Département de Physique, Université de Sherbrooke, Sherbrooke J1K2R1 Québec, Canada}
\affiliation{Canadian Institute for Advanced Research, Toronto, ON, Canada}

\author{Andreas Wallraff}
\affiliation{Department of Physics, ETH Zurich, CH-8093 Zurich, Switzerland}
\affiliation{ETH Zurich - PSI Quantum Computing Hub, Paul Scherrer Institute,CH-5232 Villigen, Switzerland}
\affiliation{Quantum Center, ETH Zurich, 8093 Zurich, Switzerland}
\author{Sebastian Krinner}
\affiliation{Department of Physics, ETH Zurich, CH-8093 Zurich, Switzerland}

\date{September 13, 2023}

\begin{abstract}

Quantum computers will require quantum error correction to reach the low error rates necessary for solving problems that surpass the capabilities of conventional computers. 
 One of the dominant errors limiting the performance of quantum error correction codes across multiple technology platforms is leakage out of the computational subspace arising from the multi-level structure of qubit implementations. Here, we present a resource-efficient universal leakage reduction unit for superconducting qubits using parametric flux modulation. This operation removes leakage down to our measurement accuracy of $7\cdot 10^{-4}$ in approximately $\ns{50}$ with a low error of $2.5(1)\cdot 10^{-3}$ on the computational subspace, thereby reaching durations and fidelities comparable to those of single-qubit gates.
 We demonstrate that using the leakage reduction unit in repeated weight-two stabilizer measurements reduces the total number of detected errors in a scalable fashion to close to what can be achieved using leakage-rejection methods which do not scale.
 Our approach does neither require additional control electronics nor on-chip components and is applicable to both auxiliary and data qubits. These benefits make our method particularly attractive for mitigating leakage in large-scale quantum error correction circuits, a crucial requirement for the practical implementation of fault-tolerant quantum computation.

\end{abstract}

\maketitle

Quantum error correction (QEC) protocols \cite{Gottesman2010, Terhal2015} offer a promising path to close the gap between physical error rates achievable on quantum computing devices and the low logical error rates necessary to solve computational problems that are intractable for conventional computers~\cite{Preskill2018}. 
However, the efficient suppression of logical errors typically relies on the assumption that physical errors occur independently both in space and time, and that physical systems used as qubits have no more than two levels~\cite{Knill1998a, Aharonov1999}. 
Yet leakage, a phenomenon in which an excitation leaves the two-level computational subspace used to perform quantum operations, is a source of highly correlated errors~\cite{McEwen2021a, Miao2022}. Consequently, leakage poses a significant challenge to achieve fault-tolerant quantum computation~\cite{Alferis2007, Fowler2013, Ghosh2015a, Suchara2015, Bultink2020, Varbanov2020, Battistel2021}. 
Leakage occurs across a wide range of technology platforms including trapped-ion systems~\cite{Stricker2020, Hayes2020}, semiconductor quantum dots~\cite{Andrews2019},  neutral atoms~\cite{Wu2022} and superconducting circuits. For superconducting circuits, leakage arises predominantly from control inaccuracies in single-qubit gate operations~\cite{Motzoi2009, Chen2016, Werninghaus2021, Lazar2022}, two-qubit gate operations~\cite{Barends2016, Rol2019, Collodo2020, Negirneac2021} and readout~\cite{Sank2016, Shillito2022, Khezri2022}. 

To mitigate the effect of leakage, so-called leakage reduction units (LRUs) have been proposed to convert leakage errors into uncorrelated errors in the computational subspace at regular intervals during the computation~\cite{Alferis2007}.
Most proposals for LRUs consist of involved teleportation circuits~\cite{Alferis2007, Fowler2013}, of auxiliary qubit reset in combination with periodic swaps between auxiliary and data qubits~\cite{Ghosh2015a}, or of dedicated filter circuits which allow for the dissipation of only the leakage state~\cite{Thorbeck2021, Ahonen2022}, all of which add overhead to quantum error correction protocols or to the device architecture.
Therefore, initial leakage-mitigation schemes for superconducting qubits~\cite{McEwen2021a, Chen2021p} focused on removing leakage using a multi-level reset operation~\cite{Magnard2018, Egger2018, Zhou2021, McEwen2021a}. However, such an operation also resets the states of the computational subspace~\cite{Fowler2012} and can therefore only be applied to auxiliary qubits at the end of an error correction cycle.
Such a scheme was recently extended to remove leakage of data qubits using an additional leakage-swap gate followed by a second auxiliary-qubit reset operation~\cite{Miao2022}.
It is only very recently that a universal LRU, a single operation which can be applied to both data and auxiliary qubits, has been demonstrated based on the proposal of Ref.~\onlinecite{Battistel2021} using a second-order microwave-activated coupling (previously used in Ref.~\onlinecite{Magnard2018} for qutrit reset) between the leakage state and the readout resonator~\cite{Marques2023b}.

In this Letter, we present a resource-efficient, fast and universal flux-activated LRU which couples the leakage state of a flux-tunable transmon qubit~\cite{Koch2007} to its readout resonator. 
The engineered coupling, resulting from a parametric qubit frequency modulation, is a first-order transition and therefore the LRU can be fast~\cite{Beaudoin2012}, reaching durations and fidelities comparable to those of single-qubit gates. 
Additionally, unlike the method used in Ref.~\onlinecite{Miao2022}, it can also be performed when the readout resonator frequency is higher than the qubit frequency, a common architectural choice~\cite{Egger2018, Andersen2020, Marques2021, Wu2021c} to avoid complications from readout-induced transitions that arise when the readout resonator frequency is lower than the qubit frequency~\cite{Khezri2022, Cohen2023}.

\begin{figure}[t] 
  \centering
  \includegraphics{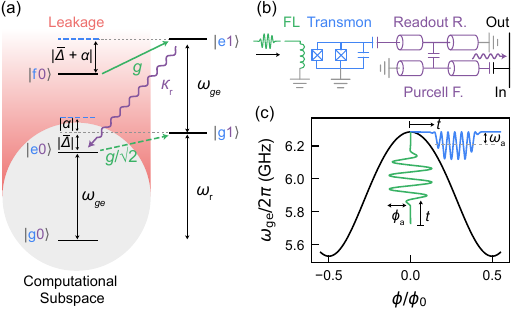}
  \caption{Concept of the leakage reduction unit. 
  (a) Energy-level diagram of a transmon qubit with transition frequency \fge{} and anharmonicity $\alpha$ coupled to a resonator mode with transition frequency \fr{}. Sidebands generated by a modulation of the qubit frequency are indicated with dashed blue lines. These sidebands enable the coupling of the leaked state $\ket{f0}$ to the state $\ket{e1}$, which decays back to the computational subspace state $\ket{e0}$. See main text for details. 
  (b) Circuit diagram of the elements required for the implementation of an LRU for a flux-tunable transmon qubit (blue): a flux line (green), and a readout resonator/Purcell-filter system (purple) coupled to a feed line (black).
  (c)  Modulating the magnetic flux in the SQUID loop of the qubit (\cref{app:device}) using a Gaussian-filtered modulation pulse $\Phi(t)$ (green line) results in a modulation of the qubit transition frequency \fge{} (blue line), leading to a parametric coupling to the resonator mode. 
}
  \label{fig:concept}
\end{figure}

\mysection{Concept}
We realize the LRU by coupling the first leakage state of a flux-tunable transmon qubit, \f, to a readout resonator-Purcell filter system which is strongly coupled to a feed line acting as a dissipative environment, as illustrated with a simplified energy-level diagram in \cref{fig:concept}(a) and a full circuit diagram of the system in \cref{fig:concept}(b). For clarity, we consider a single readout mode in the energy-level diagram although we have two hybridized readout-resonator/Purcell-filter modes~\cite{Swiadek2023}, see 
\cref{app:device} for all relevant device parameters.
We realize the coupling by applying a flux pulse $\phi(\tsymbol)$ with amplitude \modampflux{} and modulation frequency \modfreq{} to the flux line of the qubit~\cite{Strand2013,  Caldwell2018, Zhou2021}, as depicted in \cref{fig:concept}(c) and detailed in \cref{app:pulse _parameters}. 

Because the qubit is operated at its upper flux-noise-insensitive bias point (i.e.~with a DC flux bias of $\phi_\mathrm{DC} = 0$), the flux modulation results in a qubit frequency modulation with leading-order sidebands at $\pm 2\modfreq$~\cite{Beaudoin2012, Strand2013, Caldwell2018}.
The amplitude of the qubit frequency modulation is denoted \modamp, and inferred in the experiments as described in \cref{app:modulation_amplitude}.
When the high-frequency sideband [top dashed blue line in \cref{fig:concept}(a)] of $\ket{f0}$ is resonant with $\ket{e1}$, population is transferred from $\ket{f0}$ to  $\ket{e1}$. Here the first state label corresponds to the state of the transmon qubit and the second to the Fock state of the resonator mode.
This resonance condition is fulfilled when
\begin{equation} \label{eq:resonance}
     2\modfreq = |\fef[overline] -\fr| \approx |\fge[overline] + \alpha -\fr| = |\overline{\Delta} + \alpha|
\end{equation}
where $\fge$ ($\fef$) is the transition frequency from \g{} to \e{} (\e{} to \f) at the bias point, $\alpha = \fef - \fge $ is the transmon anharmonicity, and $\Delta = \fge - \fr$ is the detuning between the qubit frequency and the transition frequency $\fr{}$ of the readout resonator mode.
The overline symbol $\overline{\omega}_{kl}$ indicates the transition frequency from $\ket{k}$ to $\ket{l}$ time-averaged over the duration of the modulation pulse, i.e. $\overline{\omega}_\mathrm{ge} \approx \omega_\mathrm{ge} - \modamp$ and $\overline{\omega}_\mathrm{ef} \approx \omega_\mathrm{ef} - \modamp$.

The resulting coupling strength $g$ depends on the modulation amplitude~\cite{Beaudoin2012, Caldwell2018}
\begin{equation}
g = \sqrt{2}\gb J_1(\modamp/2\modfreq) \label{eq:g}
\end{equation}
with $\gb$ the coupling strength between the qubit and the readout resonator, and $J_1(\cdot)$ the first Bessel function of the first kind~\cite{Watson1995}.
After the leaked population has been transferred to the readout resonator, the coupled system decays back to the computational state $\ket{e0}$ on the timescale of the effective decay rate of the resonator mode $\kr/2\pi = \mhz{16.4}$ (\cref{app:device}).  

When  the flux modulation pulse is tuned to satisfy the resonance condition of the LRU [\cref{eq:resonance}], an analogous parametric transition from $\ket{e0}$ to $\ket{g1}$ with a coupling strength $g/\sqrt{2}$ is detuned by only $|\alpha|$, as illustrated in \cref{fig:concept}(a). 
It is essential to suppress residual driving of this transition because it affects the computational subspace.
To achieve this, we ensure that the bandwidth of the $\ket{e0}$ sideband is much smaller than $|\alpha|$ by filtering the rising and falling edges of the flux pulse using a Gaussian kernel of width $\sigma=\ns{5}$, see \cref{app:pulse _parameters}. Furthermore, to suppress Purcell decay of the high-frequency $\ket{e0}$ sideband, we use a device architecture with an individual Purcell filter for each qubit and readout circuit parameters which ensure that the transmission through the readout resonator-Purcell filter system is suppressed at a detuning $|\alpha|$  from resonance~\cite{Heinsoo2018, Krinner2022}.

\begin{figure}[t] 
  \centering
  \includegraphics{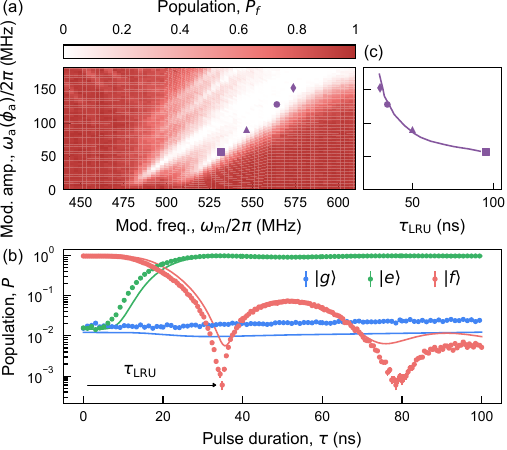}
  \caption{Calibration of the leakage reduction unit. 
  (a) \f{} population after a 100-ns-long LRU as a function of the modulation frequency \modfreq{} and the modulation amplitude \modamp. Four operating points for an LRU are indicated with purple symbols. 
  (b) Experimentally measured (dots) and simulated (lines) time evolution of the population of \g{} (blue), \e{} (green), \f{} (red) of the transmon qubit initially prepared in \f{} when applying the modulation pulse with the parameters indicated by a purple circle in (a). Error bars correspond to the standard error from 25000 single-shot measurements. (c) Measured (markers) and simulated (line) duration of the leakage reduction unit, \tlru, as a function of the modulation amplitude \modamp.}
  \label{fig:calibration}
\end{figure}

\mysection{Calibration and Characterization} \label{sec:calibration_and_characterization}
We first identify suitable operating points for the LRU, i.e.~pairs of (\modfreq, \modamp) satisfying the resonance condition.  Specifically, we prepare the qubit in \f, apply a flux modulation pulse with a fixed duration of $t=\ns{100} > 1/\kr$, and subsequently measure the transmon qubit with three-state readout~\cite{Magnard2018}. 
We sweep the modulation frequency and the modulation amplitude and identify four resonances yielding low \f{} population after \ns{100}, see the four slanted spectral lines in \cref{fig:calibration}(a). 
The high-modulation-frequency resonance doublet corresponds to the parametric transition from $\ket{f0}$ to $\ket{e1}$ from the qubit into either one of the two resonator-Purcell-filter modes.
We use the highest-frequency resonance of this doublet to implement the LRU.
The two lower-frequency resonances
are induced by a second-harmonic process, 
see \cref{app:other_resonances} for details.
For each of the spectral lines, the modulation frequency required to reach resonance increases linearly as a function of the modulation amplitude with a slope of approximately $1/2$ as the mean qubit frequency is shifted by $\modamp$ during flux modulation, see also \cref{eq:resonance}. 

In a second calibration step, we fix the modulation amplitude and frequency, and vary the duration of the pulse $\durationsymbol$ to extract the minimal duration \tlru{} of the pulse yielding the lowest population  of \f. 
For the operating point $\oppoint = (\modfreq/2\pi = \modfreqvalue ,\, \modamp/2\pi = \modampvalue)$  \mhz{} [purple circle in \cref{fig:calibration}(a)], the achieved parametric coupling $g$ is large with respect to $\kr/4$, which results in under-damped oscillations~\cite{Zhou2021} of the \f{} population with a first minimum of $6(1)\cdot 10^{-4}$ after a pulse duration of only \ns{34.5} (\ns{54.5} when accounting for the rising- and falling-edge buffers as detailed in \cref{app:pulse _parameters}), see \cref{fig:calibration}(b). 
The exhaustive depletion of the population in \f, down to the single-shot readout accuracy of approximately $7\cdot 10^{-4}$, demonstrates the high effectiveness of the LRU.
The population dynamics of all three transmon eigenstates are in good agreement with master-equation simulations [solid lines in~\cref{fig:calibration}(b)], see \cref{app:simulations_calibration} for details.

To gain further insight into the relationship between the modulation amplitude and the duration of the LRU, we measure the time evolution of the transmon population for four modulation amplitudes [purple markers in \cref{fig:calibration}(a)] and extract the corresponding \tlru. 
As expected from simulations, \tlru{} decreases approximately as $1/\modamp$ in our parameter regime, see purple markers in \cref{fig:calibration}(c).

\begin{figure}[t] 
  \centering
  \includegraphics{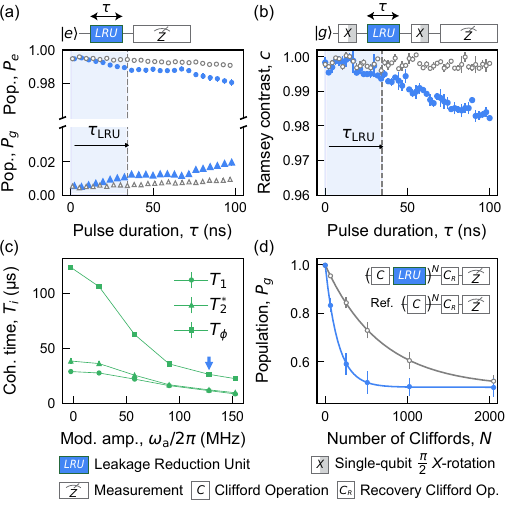}
  \caption{Effect of the LRU on the computational subspace. 
    (a) Population of the transmon states \e{} (circles) and \g{} (triangles) when inserting a flux-modulation pulse of duration $\durationsymbol$ before the measurement (blue) or a waiting time of equivalent duration (gray). The duration of an LRU for the chosen experimental parameters at the operating point \oppoint{} is indicated by the dashed gray line.
  (b) Contrast of the Ramsey fringes with (blue) and without (gray) inserting a flux modulation pulse of duration $\durationsymbol$ between the two $\pi/2$ $X$-rotations. Error bars indicate the standard errors extracted from sinusoidal fits to the data, see also \cref{app:computational_subspace}.
  (c) Effective lifetime $T_1$ (circles), dephasing time $T_2^\ast$ (triangles) and pure dephasing time $T_\phi$ (squares) as a function of the modulation amplitude \modamp. The modulation amplitude $\modamp = \mhz{\modampvalue}$ used for (a), (b) and (d) is indicated with a blue arrow. 
  (d) Interleaved randomized benchmarking (blue) performed as shown in the upper quantum circuit diagram and reference randomized benchmarking (gray) performed as shown in the lower quantum circuit diagram. 
}
  \label{fig:computational_subspace}
\end{figure}

Although leakage errors can significantly impede the performance of QEC protocols, they are infrequent, typically occurring at a rate of 0.1-1\% per qubit per QEC cycle~\cite{McEwen2021a, Krinner2022, Sundaresan2023}. Consequently, in practice the LRU acts on a state within the computational subspace most of the time, and it is therefore imperative to minimize its effect on this subspace.
We investigate the effect of the LRU on the $Z$-basis states of the computational subspace for the operating point \oppoint{} by comparing a $T_1$  measurement when applying a flux-modulation pulse of duration $\durationsymbol$ (blue), and a waiting time of equivalent duration (gray), see \cref{fig:computational_subspace}(a). We proceed analogously using a $T_2^\ast$ measurement to evaluate the effect on the $X$-basis states [\cref{fig:computational_subspace}(b)].

We observe only a small effect of the flux modulation pulse on short time scales ($\durationsymbol \leq \tlru$), with a population transfer from \e{} to \g{} of only $\sim 0.005(2)$ in the $T_1$ measurement and a similar reduction in the contrast of Ramsey fringes in the $T_2^\ast$ measurement. 
These observations suggest that, as required, the modulated pulse leaves the computational subspace mostly unaffected on the time scale of the duration of an LRU operation.

We repeat the same measurements for longer time scales to accurately extract coherence times, and find an effective lifetime of $T_1 = \mus{13.4}$ ($T_2^\ast = \mus{10.8}$) when the LRU is applied, compared to $T_1 = \mus{28.8}$ ($T_2^\ast = \mus{37.0}$) when the LRU is not applied.
 We attribute this effective reduction in lifetime to population loss due to the repeated crossing of two-level defect modes~\cite{Klimov2018, Lisenfeld2019} in the frequency spectrum of the transmon when the modulation pulse is applied (\cref{app:computational_subspace}). The reduction in $T_2^\ast$ is ascribed to the decrease in $T_1$ as well as to the decrease in the pure dephasing time $T_\phi = 2T_1T_2^\ast/ (2T_1 - T_2^\ast)$ resulting from the qubit being away from its flux-noise-insensitive bias point during the modulation pulse. 

We repeat these two characterization measurements for different operating points of the LRU and extract the effective coherence times $T_1$, $T_2^\ast$ and $T_\phi$ when applying the modulation pulse as a function of the modulation amplitude, see \cref{fig:computational_subspace}(c). For all operating points, we observe that $T_1 < T_\phi$, which indicates that errors on the computational subspace are mostly $T_1$-limited.

To extract the average error of the LRU on the computational subspace, we perform interleaved randomized benchmarking (iRB)~\cite{Magesan2012} in which the LRU is benchmarked against a perfect identity operation. For the operating point \oppoint{} [the coherence times of which are indicated by the blue arrow in \cref{fig:computational_subspace}(c)], we obtain an average gate error of 0.25(1)\%, see \cref{fig:computational_subspace}(d).
In comparison, the error for an idle operation of the same duration as the LRU is about 0.1\%,  showing that performing the LRU causes errors on the computational subspace of the same order of magnitude as coherence-limited single-qubit gates. The measured error is in good agreement with a calculated coherence limit~\cite{Assad2016} of 0.24\% which takes into account the reduction in $T_1$ and $T_2^\ast$ during the LRU.
We choose the operating point \oppoint{} for all further experiments because it provides a good compromise between LRU duration and errors on the computational subspace.

\begin{figure}[t] 
  \centering
  \includegraphics[width=1.0\columnwidth]{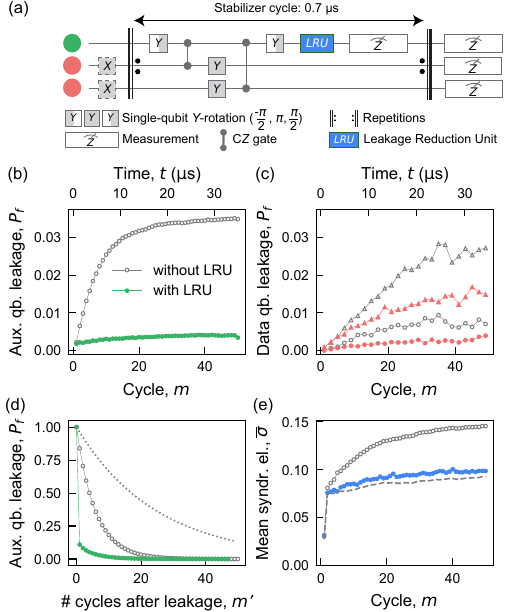}
  \caption{Integration of the LRU in a weight-two $Z$-type stabilizer measurement. 
  (a) Stabilizer circuit with one auxiliary qubit (green dot) and two data qubits (red). 
  (b) Leakage of the auxiliary qubit with (blue) and without (gray) LRU in 
 each stabilizer cycle as a function of the number of executed cycles $m$. 
 (c) Leakage of data qubits D1 (circles) and D2 (triangles) with (red) and without (gray) the LRU. 
 (d) Leakage lifetime in the stabilizer circuit with (green) and without (gray) the LRU, and in a separate characterization measurement (gray dotted line). 
 (e)~Mean syndrome element $\overline{\sigma}$ with the LRU (blue dots), with neither the LRU nor leakage-rejection (gray dots), and with leakage rejection instead of the LRU (dashed gray line).}
  \label{fig:stabilizers}
\end{figure}

\mysection{Stabilizer measurement}
To demonstrate the benefits of using an LRU in QEC experiments despite the small errors it causes on the computational subspace, we perform repeated cycles of a weight-two $Z$-type stabilizer measurement~\cite{Krinner2022} with and without LRU, see \cref{fig:stabilizers}(a) for the full quantum circuit diagram. 
The two data qubits (red dots) are initialized in one of the four $Z$-basis eigenstates and the parity of the state is mapped onto the auxiliary qubit (green dot) as shown in \cref{fig:stabilizers}(a). An LRU can be applied to the auxiliary qubit, which is subsequently measured using single-shot three-level readout~\cite{Krinner2022}. The entire stabilizer cycle of a fixed duration of \mus{0.7} is repeated $m$ times.

We find that when the LRU is applied, the accumulation of population in \f{} of the auxiliary qubit after 50 cycles, averaged over the four data-qubit input states, is reduced by approximately a factor of ten to $\sim 3.5 \cdot 10^{-3}$  [green dots in \cref{fig:stabilizers}(b)], compared to $\sim 3.4 \cdot 10^{-2}$ when the LRU is not applied (gray dots).
We observe a background residual leakage of about $2 \cdot 10^{-3}$ on average, even when we do not perform any stabilizer cycle, which is due to a frequency collision leading to state-dependent readout-induced leakage, see~\cref{app:readout_induced_leakage}. When considering solely the accumulation of leakage in addition to this background value, we calculate that the LRU leads to a 20-fold reduction in leakage accumulation.
Moreover, we find that the application of the LRU to the auxiliary qubit also reduces leakage accumulation on data qubits, as shown in \cref{fig:stabilizers}(c). We attribute this effect to a decrease in leakage transport~\cite{McEwen2021a, Miao2022} which arises only when the auxiliary qubit is in \f. The differences in leakage between the two data qubits are currently not understood.

Furthermore, we extract the effective lifetime of a leakage event in the stabilizer circuit by post-selecting on runs in which leakage is detected on the auxiliary qubit and counting the average number of cycles in which the auxiliary qubit is consecutively read out in \f{} after the initial leakage event. We find that the LRU achieves the goal of reducing the leakage lifetime on the auxiliary qubit to close to a single cycle of the repeated stabilizer measurements, while the lifetime is on the order of 6 cycles when no LRU is used, see \cref{fig:stabilizers}(d). 
In comparison, the \f-state lifetime of the auxiliary qubit in an independent $T_1$ measurement (dashed gray line) is much longer, approximately 24.6 cycles, which provides further evidence for leakage transport away from the auxiliary qubit during the repeated stabilizer measurement. 
From the reduction of the leakage lifetime, we infer that both space and time-correlated errors caused by leakage are suppressed~\cite{Miao2022}.

To further investigate the impact of the LRU on the total number of detected errors by the stabilizer, we construct the error syndrome in each cycle $\sigma_m = (1 - s_m \cdot s_{m-1}) / 2$ from the current ($m$) and the previous ($m-1$) measured stabilizer values $s$, with $\sigma=1$ indicating an error and $\sigma = 0$ indicating no error, respectively~\cite{Kelly2015, Krinner2022}.
When averaging over all circuit runs and possible data-qubit input states, we find that applying the LRU reduces the mean error syndrome value $\overline{\sigma}$ from $\sim 0.15$ to $\sim 0.1$ after 50 cycles [\cref{fig:stabilizers}(e)]. 
These results suggest that the LRU suppresses leakage-induced correlated errors and consequently reduces the total number of errors by approximately 33\%. 
To further assess the performance of our approach,
we compare the use of the LRU to a leakage-rejection  method [dashed gray line in \cref{fig:stabilizers}(e)] that discards experimental runs in which a leakage event on the auxiliary qubit is detected using three-level readout~\cite{Krinner2022, Sundaresan2023}. 
Note that this method is not suited for large-scale QEC experiments because the amount of experimental runs left after leakage-rejection decreases exponentially with the number of QEC cycles and qubits. 
By contrast, employing the LRU results in nearly the same performance as the leakage rejection method, with the key benefit of scalability.

\mysection{Discussion}
In summary, we have 
demonstrated a fast leakage reduction unit based on parametric flux modulation taking only $\sim \ns{50}$, which effectively removes leakage down to our qubit readout error of $7\cdot 10^{-4}$. Moreover, it is high-fidelity, causing only an error of $2.5(1)\cdot 10^{-3}$ on the computational subspace. Our LRU thus approaches durations and fidelities comparable to those of single-qubit gates.
We successfully integrated the LRU in a weight-two stabilizer measurement, thereby significantly improving its performance. 
Simulations show that the ability to suppress leakage will become even more relevant when executing large-scale quantum error correction circuits~\cite{Miao2022}.
In the future, the presented LRU can also be applied to data qubits and thereby further reduce the total number of errors.

The LRU introduced in this work offers several advantages compared to other recent developments in leakage suppression~\cite{Miao2022, Marques2023b}. First, our LRU is four times faster than the one presented in Ref.~\onlinecite{Marques2023b}, resulting in a reduction of idling errors on all qubits, which often constitute a substantial fraction of the total error budget~\cite{Chen2021p, Acharya2023}. 
Second, the modulation pulses are generated by the same electronics which also generates pulses for the two-qubit gates, avoiding additional cost and complexity of the experimental setup.
Finally, employing parametric coupling for realizing the LRU enables its use in a wide range of qubit-frequency configurations.
Hence, this work showcases that the flux-activated parametric LRU is a promising approach to effectively suppress leakage in large-scale error correction circuits, which is an essential requirement for the practical implementation of fault-tolerant quantum computation.

\section*{Acknowledgments}
The authors are grateful for valuable discussions with Markus M\"{u}ller. 

The team in Zurich acknowledges financial support by the Office of the Director of National Intelligence (ODNI), Intelligence Advanced Research Projects Activity (IARPA), via the U.S. Army Research Office grant W911NF-16-1-0071, by the EU Flagship on Quantum Technology H2020-FETFLAG-2018-03 project 820363 OpenSuperQ, by the National Centre of Competence in Research Quantum Science and Technology (NCCR QSIT), a research instrument of the Swiss National Science Foundation (SNSF), by the SNFS R'equip grant 206021-170731, by the EU program H2020-FETOPEN project 828826 Quromorphic and by ETH Zurich. S.K. acknowledges financial support from Fondation Jean-Jacques \& Felicia Lopez-Loreta and the ETH Zurich Foundation. The work in Sherbrooke was undertaken thanks in part to funding from NSERC, Canada First Research Excellence Fund and ARO W911NF-18-1-0411, the Ministère de l’Économie et de l’Innovation du Québec, and U.S. Department of Energy, Office of Science, National Quantum Information Science Research Centers, Quantum Systems Accelerator. M.M. acknowledges support by the U.S. Army Research Office grant W911NF-16-1-0070.
The views and conclusions contained herein are those of the authors and should not be interpreted as necessarily representing the official policies or endorsements, either expressed or implied, of the ODNI, IARPA, or the U.S. Government.

\section*{Competing interests}
The authors declare no competing interests.

\section*{Data availability}
All data is available from the corresponding authors upon reasonable request.

\appendix

\section{Device and Experimental Setup} \label{app:device}
\begin{figure}[t]
    \centering
    \includegraphics{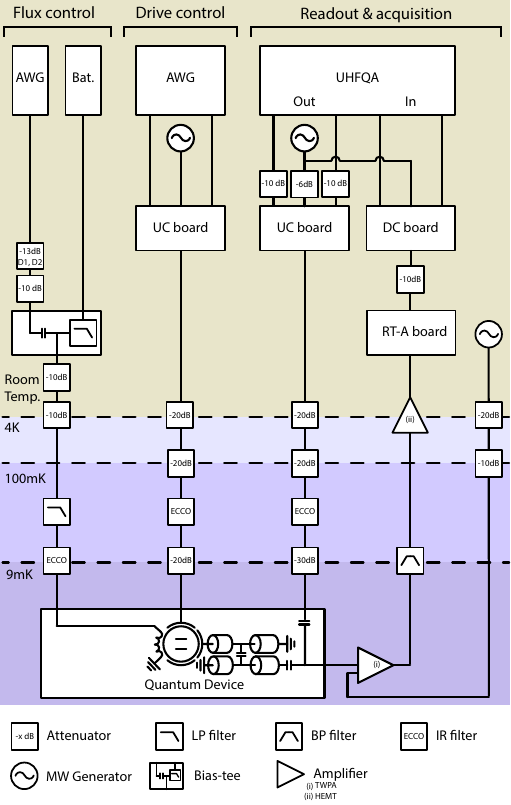}
    \caption{Schematic representation of the experimental setup. See text for details.}
    \label{fig:setup}
\end{figure}

The experiments discussed in this manuscript were conducted on a subset of three qubits of a 17-qubit device similar to the one presented in Ref.~\onlinecite{Krinner2022}, see \cref{tab:qb_params} for qubit parameters, coherence properties and error rates. 
The LRU is implemented on an auxiliary qubit A, and two neighboring data qubits D1 and D2 are used to test the integration of the LRU in repeated weight-two stabilizer measurements. 
The device is mounted at the base plate of a dilution refrigerator~\cite{Krinner2019} and connected to room-temperature control-electronics as summarized in \cref{fig:setup}.

To realize the LRU and two-qubit gates, voltage pulses generated by an AWG at a sampling rate of 2.4 GSa/s are applied to a dedicated flux line for each qubit~\cite{Krinner2022}. The pulses induce a magnetic flux in the SQUID-loop of the corresponding target qubit which controls its transition frequency. 
The pulses are predistorted to compensate for the frequency response of the flux line.  
Note that the flux line of the auxiliary qubit has 13\,dB less attenuation at the output of the AWG compared to the flux lines of the data qubits, to allow for the characterization of the LRU at large modulation amplitudes ($\modamp > \mhz{130}$).
A DC current controlling the idle frequency of the qubit is combined with the voltage pulses using a bias-tee. 

To implement single-qubit gates, an AWG generates DRAG drive pulses~\cite{Motzoi2009} at an intermediate frequency, which are then up-converted (UC) to microwave frequencies. 

We generate readout pulses using the signal-generation unit of an ultra-high frequency quantum analyzer (UHFQA). The pulses are up-converted to microwave frequencies and applied to the readout input port of the device. The transmitted signal at the output port of the device is amplified by a wideband near-quantum-limited traveling-wave parametric amplifier (TWPA) \cite{Macklin2015}, a high-electron-mobility transistor (HEMT) amplifier, and low-noise amplifiers operating at room-temperature (RT-A board)~\cite{Krinner2022}. Subsequently, the amplified signal is down-converted with an IQ-mixer in a down-conversion (DC) board, and then both demodulated and integrated using the FPGA-based acquisition unit of the UHFQA.

\begin{table}[]
    \centering
    \caption{Parameters, coherence properties and error rates of the auxiliary qubit (A) and the two data qubits (D1, D2) used in the experiment. RO stands for readout.}
\begin{tabular}{lrrr}
\toprule
Parameter &       D1 &       A &       D2 \\
\midrule
Qubit idle frequency, $\omega_\mathrm{ge}/2\pi$ (GHz)					        	    & 5.041 & 6.281 & 4.999 \\
Qubit anharmonicity, $\alpha/2\pi$ (MHz)							        & -167 & -154 & -167 \\
Lifetime, $T_1$ (\si{\micro s})							                    & 29 & 21 & 64 \\
Ramsey decay time, $T_2^*$  (\si{\micro s})				        	        & 39 & 34 & 79 \\
Echo decay time, $T_2^\mathrm{e}$   (\si{\micro s})		        		    & 51 & 35 & 99  \\
Readout frequency, $\omega_\mathrm{RO}/2\pi$ (GHz)				        	& 6.816 & 7.129 & 6.667 \\ 
Qb. freq. during RO, $\omega_\mathrm{ge}'/2\pi$ (GHz)					            	& 5.054 & 6.281 & 5.572  \\ 
Dispersive shift, $\chi/2\pi$ (MHz)			        	                    & -3.9 & -5.0 & -4.0 \\ 
RO Resonator linewidth, $\kr/2\pi$ (MHz)	                    & 6.3 & 16.4 & 8.2  \\  
Purcell filter linewidth, $\kp/2\pi$ (MHz)	                    & 26.5 & 35 & 27.6  \\ 
Purcell filter frequency, $\fp/2\pi$ (GHz)	                    & 6.855 & 7.111 & 6.689  \\ 
Qubit-RO res. coupling, $\gb/2\pi$ (MHz)	                    & 175 & 120 & 167 \\ 
RO res.- Purcell f. coupling, $J/2\pi$ (MHz)                            & 22.0 & 28.8 & 22.2\\
Two-state readout error, $\epsilon_{\mathrm{RO}}^{(2)}$ (\%)			    & 0.56 & 0.54 & 0.52  \\
Three-state readout error, $\epsilon_{\mathrm{RO}}^{(3)}$ (\%)			    & 1.29 & 1.53 & 1.53  \\ 
Two-qubit gate error, $\epsilon_\mathrm{2Q}$ (\%)   & \multicolumn{3}{c}{\,\,\,\,0.82 \,\,\, 0.73}  \\
A-D1 gate leakage rate, $\lambda_{\rm D1, A}$ (\%)  & 0.01\;\, & 0.03\;\, & n/a \\
A-D2 gate leakage rate, $\lambda_{\rm A, D2}$ (\%) & n/a\;\, & 0.03\;\, & 0.00 \\
\bottomrule
\end{tabular}
    \label{tab:qb_params}
\end{table}

\section{LRU Pulse Parameters} \label{app:pulse _parameters}
\begin{figure}[t]
    \centering
    \includegraphics{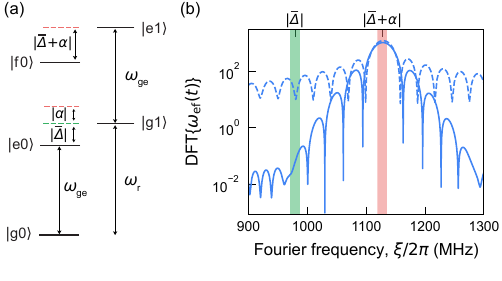}
    \caption{Effect of Gaussian filtering on the frequency spectrum of the qubit transition-frequency sidebands when applying the modulation pulse. (a) Energy-level diagram of the transmon-resonator system. The red dashed lines indicate the high-frequency first-order sidebands, detuned by $|\overline{ \Delta} + \alpha| $ from $\ket{e0}$ and $\ket{f0}$. 
    The green dashed line indicates the frequency which drives the undesired transition from $\ket{e0}$ to $\ket{g1}$, detuned by $|\overline{\Delta}|\approx \mhz{979}$ from $\ket{e0}$. 
    (b) Frequency spectrum of the first order sideband of $\ket{f0}$ with (solid blue line) and without (dashed blue line) filtering the modulation pulse using a Gaussian kernel of width $\sigma = \ns{5}$ at the operation point \oppoint{} described in the main text. The red shaded (green shaded) area indicates the overlap between the first order sideband and the linewidth $\sim \kr/2\pi$ of $\ket{e1}$ ($\ket{g1}$) at the operating point \oppoint. The spectrum is obtained by applying a discrete Fourier transform (DFT) to the calculated ef-transition frequency of the qubit when applying the flux modulation pulse, $\fef(t)$, and taking the modulus thereof.}
    \label{fig:pulse_shaping}
\end{figure}

When the LRU is enabled, which occurs when the modulation frequency is chosen such that the high-frequency sideband of $\ket{f0}$ is resonant with $\ket{e1}$, the transition between $\ket{e0}$ and $\ket{g1}$  is detuned by the anharmonicity $|\alpha|$, as illustrated by the energy-level diagram in \cref{fig:pulse_shaping}(a). 
To suppress the off-resonant drive of this undesired transition which affects the computational subspace, we filter the flux pulse by convolving the square-shaped modulation pulse with a Gaussian kernel (characterized by its width $\sigma$). This filtering process reduces the bandwidth of the sidebands, see \cref{fig:pulse_shaping}(b) for a comparison between the spectrum of the first-order sideband of $\ket{f0}$ for $\sigma=\ns{5}$ (solid blue line) and without filtering, i.e.~$\sigma = \ns{0}$ (dashed blue line). 
Consequently, the undesired overlap between the $\ket{e0}$ sideband and $\ket{g1}$ [green shaded area in \cref{fig:pulse_shaping}(b)], is suppressed by several orders of magnitude compared to when no filtering is applied.

The resulting voltage pulse applied to the input of the flux line is given by 
\begin{equation}\label{eq:voltage_flux}
    v(\rtime)=\frac{V_\mathrm{a}}{2}\cos(\omega_\mathrm{m}\rtime)\left[\mathrm{erf}\left(\frac{\rtime - \bufferduration}{\sqrt{2}\sigma}\right)-\mathrm{erf}\left(\frac{\rtime- \bufferduration - \durationsymbol}{\sqrt{2}\sigma}\right)\right]
\end{equation}
where $V_\mathrm{a}$ is the voltage amplitude of the pulse, $\bufferduration$ is the duration of the start and end buffer, $\mathrm{erf(\cdot)}$ is the Gaussian error function, $\durationsymbol$ is the duration of the square-shaped modulation pulse and $\rtime\in[0, \durationsymbol + 2\cdot \bufferduration]$ is the time.

The pulse duration required to reach the first minimum in \f{} population [see \cref{fig:calibration}(c) in the main text for an example], denoted as $\tlru$, increases with $\sigma$ due to the impact of $\sigma$ on the shape of the rising and falling edges of the pulse. Indeed, a larger $\sigma$ reduces the effective duration at which the modulation amplitude is maximal. 
Furthermore, to avoid significant clipping of the waveform, the rising- and falling-edge buffers are adapted such that $\bufferduration = 2\sigma$, which also results in a longer total LRU duration for larger $\sigma$. 

\begin{figure}[t]
    \centering
    \includegraphics{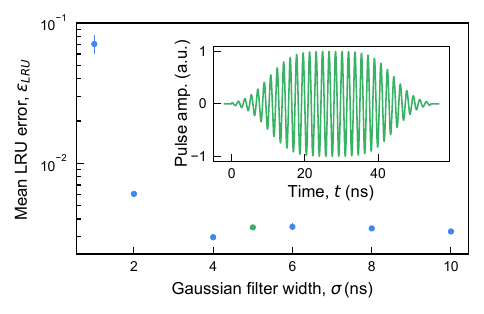}
    \caption{Mean error of the LRU benchmarked against the identity gate with interleaved randomized benchmarking for different Gaussian filter kernel widths $\sigma$ at the operating point \oppoint. The green dot corresponds to the chosen $\sigma$ for the experiments in the main text. The inset shows the voltage pulse used for the stabilizer measurements presented in the main text, i.e.~with a pulse duration of $\tlru = \ns{34.5}$, $\sigma=\ns{5}$, and rising- and falling edge buffers of $\bufferduration = 2\sigma = \ns{10}$.  Error bars correspond to the standard deviation of errors determined in 3 consecutive runs.}
    \label{fig:RB_sigmas}
\end{figure}

To investigate the tradeoff between LRU duration and Gaussian filter kernel width, we use interleaved randomized benchmarking (as described in the main text) and compare LRUs for different values of $\sigma$ at the LRU operating point \oppoint.
We find that the LRU error is high for very small $\sigma$ because the bandwidth of the qubit frequency modulation starts to overlap with the $\ket{e0}$ to $\ket{g1}$ transition, see \cref{fig:RB_sigmas}. For $\sigma\gtrsim \ns{4}$, the error saturates around $3.5 \cdot 10^{-3}$, which we mostly attribute to decoherence during the LRU and interactions with two-level defects (\cref{app:computational_subspace}). 
We choose $\sigma=\SI{5}{\nano\second}$ and $\bufferduration = 2\sigma = \ns{10}$ to ensure negligible driving of undesired transitions, but maintaining a fast LRU. 
Note that the LRU benchmarked in this measurement  showed slightly worse performance than the one presented in the main text.

\section{Determination of the modulation amplitude}
\label{app:modulation_amplitude}

\begin{figure}[t]
    \centering
    \includegraphics{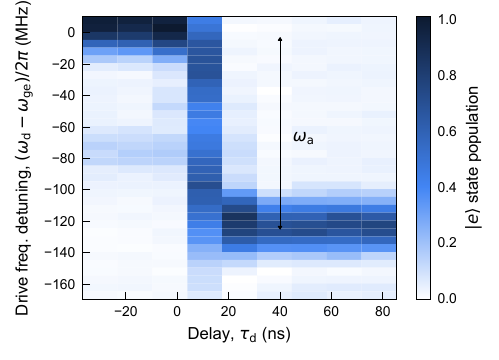}
    \caption{Excited state population as a function of the time delay between the center of the $\pi$-pulse and the rising edge of the flux pulse, $\tau_\mathrm{d}$, and the detuning between the frequency of the $\pi$-pulse $\omega_\mathrm{d}$ and the idle frequency $\fge$ of the qubit. The arrow indicates the extracted modulation amplitude \modamp.}
    \label{fig:flux_pulse_scope}
\end{figure}

To accurately determine the modulation amplitude \modamp{} at the operating point \oppoint, we measure the \e{} population of the auxiliary qubit A after applying a $\pi$-pulse during a 1-µs-long modulated flux pulse with $\modfreq/2\pi = \mhz{\modfreqvalue}$. 
Because the qubit is excited only when the drive tone is resonant with the qubit frequency, sweeping the frequency of the $\pi$-pulse, $\omega_\mathrm{d}$, as well as its relative timing to the flux pulse, $\tau_\mathrm{d}$, allows to determine the qubit frequency 
during the flux pulse~\cite{Hellings2022}. 
Note that the time-resolution of this method is limited by the duration of the drive pulse which has a Gaussian envelope with a width of $\sigma = \ns{10}$. Consequently, the fast oscillation of the qubit frequency resulting from the flux modulation [$1/(2\cdot \mhz{\modfreqvalue})~\approx~\ns{0.9}$ at the operating point \oppoint] cannot be resolved. Rather, this approach provides a good estimate of the average qubit frequency over the time window of the duration of the drive pulse.  We extract the modulation amplitude by comparing the frequency which excites the qubit well before the flux pulse ($\tau_\mathrm{d} \ll \ns{0}$) i.e.~the idle frequency, and the frequency which excites the qubit after the rising edge of the flux pulse ($\tau_\mathrm{d} > \ns{30}$), and find a mean qubit frequency during modulation $\overline{\omega}_\mathrm{ge}/2\pi\approx\SI{6.153}{\giga\hertz}$ which yields $\modamp/2\pi\approx\SI{128}{\mega\hertz}$ for the operating point \oppoint, see \cref{fig:flux_pulse_scope}. 

Because of the frequency-dependent attenuation of the flux line, performing explicit measurements of the modulation amplitude \modamp{}, as described above, for a two-dimensional sweep of the flux pulse modulation frequency \modfreq{} and amplitude $\modampflux{}$ would be a tedious endeavor. To gain a qualitative understanding of the modulation amplitude dependence on the modulation frequency, as presented in \cref{fig:calibration}(b), 
we therefore use a simpler yet less accurate approach. 
Specifically, we calculate the modulation amplitude based on the expected flux amplitude at the SQUID loop, taking into account the frequency response of the flux line and the non-linear qubit-frequency dependence on flux [black line in \cref{fig:concept}(c)].
Using this method, we obtain an estimate of the modulation amplitude of $\modamp/2\pi \approx \mhz{112}$ for the operating point \oppoint, i.e.~within $\sim\SI{15}{\percent}$ of the value extracted using the precise characterization measurement described above.
Subsequently, we adjust all modulation amplitude estimates in the landscape of \cref{fig:calibration}(b) by rescaling them linearly based on the calibration point derived from the precise measurement of the modulation amplitude at the operating point \oppoint.

\section{Other Parametric Resonances} \label{app:other_resonances}
\begin{figure}
    \centering
    \includegraphics{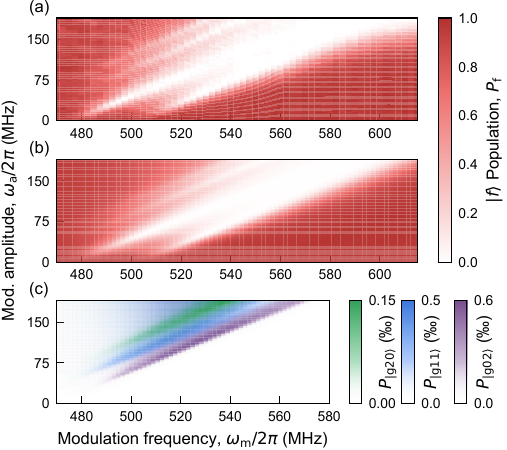}
    \caption{Measured (a) and simulated (b) \f{} population after a 100-ns-long LRU as a function of the modulation frequency and the modulation amplitude. (c) Simulated population of states $\ket{g20}$ (green), $\ket{g11}$ (blue), and $\ket{g02}$ (purple) after a 100-ns-long LRU.}
    \label{fig:landscape-simulation}
\end{figure}

We observe four main resonances which deplete the \f{}  population of the transmon qubit for the modulation frequencies and modulation amplitudes swept in \cref{fig:calibration}(a) of the main text, also reproduced in \cref{fig:landscape-simulation}(a) for comparison to the master-equation simulations presented in \cref{fig:landscape-simulation}(b). 
The two right-most resonances correspond to parametric transitions from $\ket{f00}$ to $\ket{e10}$ and $\ket{e01}$, where the first state label describes the transmon state and the second and third denote the Fock state occupation numbers of the lower- and higher-frequency mode of the hybridized Purcell filter/readout resonator system, respectively. 
The mode hybridization arises in the parameter regime of our readout architecture~\cite{Swiadek2023}, in which $J\gtrsim \kp$ and $|\Delta_\mathrm{R,P}| \ll J$. Here, $J$ represents the coupling strength between the readout resonator and the Purcell filter, $\kp$ denotes the linewidth of the Purcell filter, and $|\Delta_\mathrm{R,P}|$ is the detuning between the Purcell filter and the bare readout resonator frequency.
The two resonances in \cref{fig:landscape-simulation}(a) are separated in flux modulation frequency by about \mhz{26}, or \mhz{52} in qubit frequency due to the qubit being modulated at its upper flux-noise-insensitive bias point, which is in good agreement with the expected splitting of the hybridized modes of about $2J/2\pi = \mhz{57.5}$ and master-equation simulations of the full system, see \cref{fig:landscape-simulation}(b). While either of the two modes can be used to realize the LRU, we choose the mode with the highest frequency [right-most slanted line in \cref{fig:landscape-simulation}(a)] as it is further detuned from the closest parametric transition affecting the computational subspace, i.e.~the transition from $\ket{e00}$ to $\ket{g01}$.

We observe two additional resonances at lower modulation frequencies in \cref{fig:landscape-simulation}(a), which we attribute to transitions from $\ket{f00}$ to $\ket{g20}$, and from $\ket{f00}$ to $\ket{g11}$ driven via their second harmonic. Indeed, master-equation simulations confirm that $\ket{g20}$ and $\ket{g11}$ are residually populated after the 100-ns-long LRU for similar combinations of (\modfreq, \modamp), as indicated by the green and blue slanted lines in \cref{fig:landscape-simulation}(c), respectively.
The transition from $\ket{f00}$ to $\ket{g02}$ [the population of which is shown in purple shades in \cref{fig:landscape-simulation}(c)] is only barely visible in the experimental data as it is close to the frequency of the transition from $\ket{f00}$ to $\ket{e10}$.
While these transitions could in principle also be used to realize the LRU, they would result in a considerably longer LRU duration for a fixed modulation amplitude, because the parametric coupling achieved using the second harmonic is weaker than the one obtained using the first harmonic~\cite{Caldwell2018}. In addition, the modulation frequencies enabling these transitions are closer to the modulation frequency enabling the undesired transitions from $\ket{e00}$ to $\ket{g01}$ or $\ket{g10}$.

\section{Master Equation Simulations}
\label{app:simulations_calibration}
In this Appendix, we provide a concise description of the numerical master-equation simulations employed to support the analysis of the experiment. The system consists of a transmon with charge and phase operators $\hnt$ and $\hphit$ respectively. It has a charging energy $E_C/2\pi = \mhz{159}$ and a flux-tunable Josephson energy $E_J(\phi(t))$ where $\phi(t)$ is the external flux. The transmon is coupled to the readout resonator whose annihilation operator is $\ha$ with a bare frequency $\frb$. The readout resonator itself is coupled to a Purcell filter with frequency $\fp$ and mode annihilation operator $\hf$. The time-dependent Hamiltonian thus takes the form 
\begin{align}\nonumber
        \hat{H}(\rtime)/\hbar
        &=
        4 E_C \hat{n}_\mathrm{t}^2
        - E_J(\phi(\rtime)) \cos \hphit
        + i \gb{}_{,c}(\ha-\ha^\dagger)\hnt
        \\ \label{app_D:Hamiltonian}
        &
        +
        \frb \ha^\dagger \ha 
        + \fp \hf^\dagger \hf
        -J (\ha - \ha^\dagger)(\hf - \hf^\dagger).
    \end{align}
Here,  $\gb{}_{,c}$ is the coupling between the transmon's charge operator 
and the readout resonator, and $J$ is the coupling between Purcell filter and readout resonator respectively. Note that although one could safely make the rotating-wave approximation to the Hamiltonian \cref{app_D:Hamiltonian} eliminating terms like $-J (\ha \hf +\rm{h.c.})$, we include them in the numerical simulations. Furthermore, we keep six transmon eigenstates and 6 photons in both the resonator and filter modes.

To model dissipation, we assume that there is only dissipation on the (bare) Purcell filter mode. The master equation to be simulated is then
    \begin{align}\label{app_D:ME}
        \partial_\rtime \hrho= -\frac{i}{\hbar}[\hH(\rtime), \hrho] + \kp \mathcal{D}[\hf]\hrho
    \end{align}
where $\kp$ is the bare dissipation rate. Note that here we have written the master equation in the bare basis: the decay rate $\kr$ in the main text thus corresponds to the Purcell decay rate which stems from the strong hybridization of the bare resonator and filter modes. 

Since the couplings $\gb{}_{,c}$ and $J$ dress the transmon, resonator and filter, it is crucial in the simulations that we make the distinction between the bare and dressed modes. The time-dependent-flux further complicates this picture. Therefore, we  split the transmon Hamiltonian into two parts, namely the time-independent and time-dependent part of the Hamiltonian~\cite{Koch2007}
\begin{align}
   4 E_C \hnt^2 - E_J(\phi(\rtime)) \cos \hphit 
   \equiv
   \hH_\mathrm{t} + \delta E_J(\rtime) \cos \hphit
\end{align}
where $\hH_\mathrm{t} = 4E_c \hat{n}_\mathrm{t}^2-\ejmax\cos \hphit$ is the standard transmon Hamiltonian and 
\begin{align}
    \delta E_J(\rtime)
    \equiv
    \ejmax
    \left( 
      1-|\cos \phi(\rtime)|
    \sqrt{1 + d^2 \tan ^2 \phi(\rtime)}
    \right)
\end{align}
with $\ejmax /2\pi=33.094\,$GHz the Josephson energy at the upper first-order insensitive flux bias point and $d=0.776$ the junction asymmetry \cite{Blais2021}. Note that this picture is slightly different than that presented in the main text, where for instance $g$ in \cref{eq:g} is implicitly defined in the rotating frame of the instantaneous Hamiltonian. However, both frames coincide during initialization and measurements where the time-dependent external fluxes are set to zero, and thus the dressed eigenstates in both pictures are the same. 

With this separation,  we can now define the dressed modes, which constitute the eigenstates of the Hamiltonian Eq.~(\ref{app_D:Hamiltonian}) in the absence of flux drive (i.e $\phi(\rtime)=0$). We let $|i_\mathrm{t}, j_\mathrm{r}, k_\mathrm{f} \rangle$ denote the product state in the bare basis where $i_t$ labels the $i$-th eigenstate of the transmon and $j_\mathrm{r}$, $k_\mathrm{f}$ index the Fock states of resonator and Purcell filter respectively. We then denote the dressed basis by their tilded counterparts $|\tilde{i}_\mathrm{t}, \tilde{j}_\mathrm{r}, \tilde{k}_\mathrm{f} \rangle $.
Note that due the dispersive regime considered here, we can safely identify which eigenstates are mostly transmon-like. Since population measurements do not discriminate between how many photons are in the readout resonator or Purcell filter modes, we must trace over all such states. Thus, for instance, when we plot the population in the ground state numerically we are computing
    \begin{align}
        P_g(\rtime)
        \equiv
        \trace
        \left(
        \left[
        \sum_{\tilde{j}_\mathrm{r}, \tilde{k}_\mathrm{f}}
        |\tilde{g}, \tilde{j}_\mathrm{r}, \tilde{k}_\mathrm{f} \rangle
        \langle
        \tilde{g}, \tilde{j}_\mathrm{r}, \tilde{k}_\mathrm{f}
        |
        \right]
        \hat{\rho}(\rtime)
        \right)
    \end{align}
    where the sum over $\tilde{j}_\mathrm{r}$ and $\tilde{k}_\mathrm{f}$ performs a trace over the bosonic modes.

    With the Hamiltonian \cref{app_D:Hamiltonian}, master equation \cref{app_D:ME} and state-identification procedure presented above, one can reproduce the plots in \cref{fig:calibration} assuming one is given the functional form of the external flux $\phi(\rtime)$. The flux will be proportional to the applied voltage $v(\rtime)$ [\cref{eq:voltage_flux}], 
        \begin{equation}
         \phi(\rtime)
     =
     \frac{D}{V_\mathrm{a}}v(\rtime),
        \end{equation}
where $D$ takes into account the constant of proportionality between the applied voltage and corresponding flux. To get the value of $D$, we 
diagonalize the time dependent Hamiltonian $\hat{H}(\rtime)$ \cref{app_D:Hamiltonian} to obtain the instantaneous (dressed) qubit frequency $\fge(\rtime)$ by analogy to \cref{fig:flux_pulse_scope}. The numerical value of $D\approx 2\times 0.34 \: \phi_0/\pi$ is chosen to reproduce the experimental modulation depth $\modamp/2\pi \approx 128$ MHz at the operating point \oppoint, that is  $\fge^{\mathrm{max}}-\Bar{\omega}_{\mathrm{ge}}\approx \modamp$, where $\fge^{\mathrm{max}}$ and $\Bar{\omega}_{\mathrm{ge}}$ are the maximum qubit transition frequency and the average value of the instantaneous qubit frequency $\fge(\rtime)$, respectively.
The Hamiltonian parameters of \cref{app_D:Hamiltonian} are obtained from a simultaneous fit of the qubit transition frequency as a function of flux,  the readout resonator-Purcell filter spectroscopy as a function of flux, and the residual \f{} population of the transmon at the first minimum and first maximum of the time evolution of the system [see \cref{fig:calibration}(b)]. 
Using this procedure results in parameters which yield a time evolution well-describing the experimental data presented in \cref{fig:calibration}(b). The calculated parameters are in reasonable agreement with independent spectroscopic measurements.

\section{Effect of the LRU on the Computational Subspace} \label{app:computational_subspace}
\begin{figure}[t]
    \centering
    \includegraphics{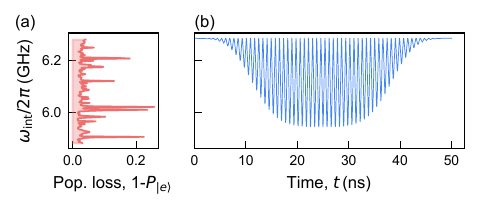}
    \caption{Weakly coupled defects crossed by the qubit transition-frequency during the LRU. (a) Population loss of the qubit initially prepared in \e{} and then pulsed to a fixed frequency for \SI{1}{\micro\second}. (b) Instantaneous qubit frequency during the LRU at the operating point \oppoint{} and with a pulse duration of $\tlru = \ns{34.5}$ filtered by a Gaussian kernel of width $\sigma=\ns{5}$ and truncated at $2\sigma=\ns{10}$ on each side, adding up to a total duration of \ns{54.5}  (as used in the experiments in the main text).}
    \label{fig:defect_crossing}
\end{figure}

During the modulated flux pulse, the qubit frequency rapidly oscillates between the idle frequency $\fge$ and $\fge - 2\modamp$, which increases the possibility of interacting with weakly coupled two-level defects in this frequency interval.
We characterize the defect mode spectrum using an independent measurement in which we tune the qubit frequency to $\omega_\mathrm{int}$ for \SI{1}{\micro\second} after a $\pi$-pulse, as described in Ref.~\cite{Krinner2022}, and show the resulting population loss in \cref{fig:defect_crossing}(a).  
We find that the calculated instantaneous qubit frequency during the LRU at the operating point \oppoint{} [\cref{fig:defect_crossing}(b)] repeatedly crosses several weakly coupled defects during the operation of the LRU, and hypothesize that the reduction in lifetime of the \e{} state described in the main text primarily arises from the interaction with these defects. 

The modulation of the qubit transition frequency also induces a coherent phase rotation of the qubit, which we counteract by applying a virtual $Z$-gate of equal magnitude and opposite sign. Note that applying this phase correction is not required when the LRU is applied to auxiliary qubits in stabilizer measurements, because the auxiliary qubits are in eigenstates of the $Z$-basis when the LRU is applied. Conversely, the phase correction is necessary when the  LRU is applied to data qubits or used in a randomized benchmarking sequence.

\begin{figure}[t]
    \centering
    \includegraphics{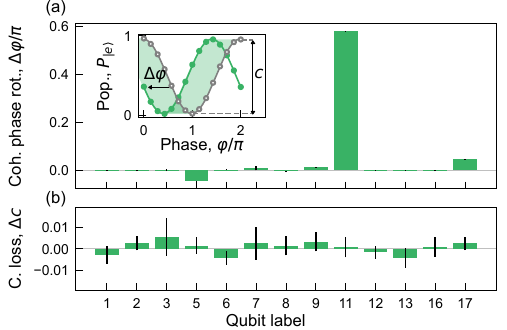}
    \caption{Effect of a 1-µs-long modulated flux pulse applied to qubit A on the other 13 functional qubits of the device. (a) Coherent phase rotation angle $\Delta\varphi$ caused by the modulated pulse. We extract the phase rotation by comparing two Ramsey measurements in which we sweep the phase of the second $\pi$-half pulse to measure the phase of the qubit without (gray circles in inset) and with (green dots in inset) applying the modulated pulse. The inset depicts the measured (dots) and fitted (lines) phase of qubit D2. (b) Loss of contrast $\Delta c$ of the Ramsey fringes when applying the modulated pulse, where the contrast $c$ corresponds to the peak-to-peak amplitude of the cosine as depicted in the inset of (a). Error bars (black lines) indicate the standard errors extracted from sinusoidal fits to the data. }
    \label{fig:crosstalk}
\end{figure}

In addition to characterizing the impact of the flux modulation pulse on the computational subspace of the targeted qubit, it is crucial to evaluate its impact on the computational subspace of other qubits. 
Flux crosstalk may induce frequency deviations of neighboring qubits, leading to coherent phase rotation and/or dephasing. To investigate this effect, we repeat the phase-sensitive measurement described in the main text, but measure the phase of neighboring qubits instead of qubit $A$ whose flux is modulated.  For each qubit $i \neq$ A, we compare the phase with and without applying the flux pulse on qubit A to extract the coherent phase rotation $\Delta\varphi$ induced by the pulse and the loss of contrast of the Ramsey fringes $\Delta c$, see \cref{fig:crosstalk}. We apply a flux modulation pulse at the operating point \oppoint{} described in the main text with a duration of \mus{1} (i.e. approximately 30 times longer than the duration of the LRU at the operating point \oppoint) to amplify the resulting errors.

Our results indicate that for a 1-µs-long modulated flux pulse at the operating point \oppoint, all measured qubits, except qubit D2, experience a coherent phase rotation of $|\Delta\varphi|/\pi \leq 0.05$,  see \cref{fig:crosstalk}(a). This corresponds to $|\Delta\varphi|/\pi \leq 0.0016$, i.e.~$|\Delta\varphi| \leq  0.3\,^{\circ}$, for the duration of a single LRU operation. 
We also find that the contrast loss is close or equal to zero within error bars [\cref{fig:crosstalk}(b)] for all measured qubits, indicating that the dephasing of neighboring qubits caused by flux crosstalk is negligible.
These findings suggest that the impact of flux crosstalk is generally small. 

We attribute the larger phase rotation on qubit D2, $\Delta\varphi \approx 3.6^{\circ}$ for the duration of an LRU operation, to the larger flux crosstalk between qubit A and qubit D2. Indeed, the cross-flux ratio~\cite{Krinner2022} $\log_{10}\;\frac{|\mathrm{d}\Phi_j/\mathrm{d}V_{A}|}{\mathrm{d}\Phi_{A}/\mathrm{d}V_{A}} \approx -2$ for $j =$ D2, but is on the order of -3 or smaller for all other qubits. Here, ${\rm d}\Phi_j/{\rm d}V_A$ is the sensitivity of the flux generated at the SQUID loop of qubit $j$ when varying the voltage applied to the flux line of qubit A, see Ref.~\cite{Krinner2022} for details. 
When integrating the LRU in surface code experiments, we intend to correct crosstalk-induced coherent phase rotations by utilizing virtual-$Z$ rotations of equal magnitude and opposite sign.
However, we did not employ this approach in the repeated stabilizer measurement presented in the main text since qubit D2 remains in the eigenstate of the $Z$-basis throughout the execution of the quantum circuit.

\section{State-dependent Readout-induced Leakage} \label{app:readout_induced_leakage}
\begin{figure}[H]
    \centering    \includegraphics[width=0.45\textwidth]{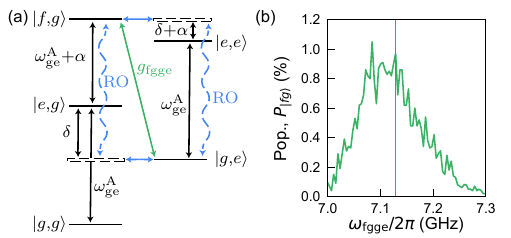}
    \caption{State-dependent readout-induced leakage. (a) Energy-level diagram of qubit A with qubit D1 in $\ket{g}$ (left) and $\ket{e}$ (right). Virtual states are shown by dashed lines. The curved arrow indicates the driving of transitions enabled by the readout tone. 
    (b) Measured $\ket{fg}$ population when preparing qubits A and D1 in $\ket{ge}$ and sweeping the data qubit frequency. The x-axis corresponds to the calculated detuning of $\ket{fg}$ and $\ket{ge}$ during the readout. The vertical blue line indicates the readout frequency of the auxiliary qubit.}
    \label{fig:fgge}
\end{figure}

The residual leakage in the stabilizer measurement is higher than the \f{} level population of $6(1)\cdot 10^{-4}$ achievable with the LRU in an isolated setting, see \cref{fig:calibration}(c). 
We believe the dominant factor limiting the residual population is a readout-induced two-qubit transition from $\ket{ge}$ to $\ket{fg}$, where the first and second state labels denote the state of the auxiliary and data qubit, respectively. This transition has been used in prior work~\cite{Krinner2020a} to perform two-qubit gates by applying a tone corresponding to the frequency detuning between these two states to the drive line of one of the qubits.
However, because the readout resonator is strongly coupled to the qubit, this transition can also be activated by a drive tone applied to the readout line.
In our configuration, the readout frequency of qubit A coincides with the ac-Stark shifted transition frequency from $\ket{ge}$ to $\ket{fg}$ during the readout. Specifically, the readout frequency of qubit A is $\SI{7.129}{\giga\hertz}$ and the transition frequency during the readout varies between $\SI{6.846}{\giga\hertz}$ and $\SI{7.535}{\giga\hertz}$ for qubit A \& D1 and $\SI{6.778}{\giga\hertz}$ and $\SI{7.577}{\giga\hertz}$ for qubit A \& D2 due to the ac-Stark shift. An energy-level diagram of the two-qubit system illustrates the drive of the transition through virtual states [see \cref{fig:fgge}(a)].

To observe the transition from $\ket{ge}$ to $\ket{fg}$ in an independent measurement, we prepare the system in $\ket{ge}$, apply the readout tones to both qubits and infer the $\ket{fg}$ population after the readout.
We reduce the readout amplitude and hence the ac-Stark shift of the auxiliary qubit compared to the other experiments presented in this work so that the ac-Stark shifted transition frequency is above the readout tone of the auxiliary qubit $\omega_\mathrm{fgge}>\omega_\mathrm{ro}$. 
To maintain high single-shot readout fidelity on the auxiliary qubit, we increase the duration of the readout tone from \ns{250} to \ns{600}. 
We sweep the transition-frequency of the data qubit during the readout using a flat-top flux pulse, thus sweeping $\omega_\mathrm{fgge}$ through the readout frequency of the auxiliary qubit. We observe an increased transition probability as $\omega_\mathrm{fgge}$ becomes resonant with the fixed-frequency readout-tone of the auxiliary qubit  [\cref{fig:fgge}(b)], indicating a drive of the two-qubit transition through the readout-tone and confirming state-dependent leakage induced by the readout tone of the auxiliary qubit.

In the stabilizer measurement, the described transition induces leakage of the auxiliary qubit if the auxiliary and data qubit are in the $\ket{ge}$ state before readout. The worst-case scenario occurs when both data qubits are in $\ket{e}$ and the auxiliary qubit is in \g{} at the end of the stabilizer cycle because both readout-induced transitions are driven. 
This results in a residual leakage population of about 0.37(2)\% for the auxiliary qubit after a single stabilizer cycle. Conversely, when both data qubits are in the ground state at the end of the cycle, neither of the readout-induced leakage transitions occur. In this case, we measure a leakage population of only about 0.04(5)\% for the auxiliary qubit, which is consistent with the value of of $0.06(1)\%$ shown in \cref{fig:calibration}(c).

We note that the observed state-dependent readout induced leakage is specific to the frequency configuration of data qubits, auxiliary qubits and readout resonators on our device. Hence, this source of leakage can be avoided in future experiments by adjusting the design frequencies.

\newpage

\end{document}